\documentclass[twocolumn,floatfix,altaffilletter,preprintnumbers,superscriptaddress,tightenlines,showpacs,showkeys,nofootinbib]{revtex4-1}

\usepackage[dvips]{graphicx}

\usepackage{siunitx}
\usepackage{array,amsmath,amsthm,feynmf,mathtools}
\usepackage{mathtools}
\usepackage{epsfig}
\usepackage{graphicx}
\usepackage{color}
\usepackage{slashed}
\usepackage{amssymb}
\usepackage{enumitem}
\usepackage{amsmath}
\usepackage{hyperref}
\usepackage{breakurl}
\usepackage{xcolor}

\makeatletter

\def\be{\begin{equation}}
\def\ee{\end{equation}}
\newcommand{\bea}{\begin{eqnarray}}
\newcommand{\eea}{\end{eqnarray}}

\begin{document}

\preprint{DESY-22-125}

\title{Bouncing Dark Matter}

\author{Lucas Puetter}
\affiliation{II. Institute of Theoretical Physics, Universität Hamburg, 22761 Hamburg, Germany}

\author{Joshua T.~Ruderman}
\affiliation{Center for Cosmology and Particle Physics, Department of Physics,
New York University, New York, NY 10003, USA}
\author{Ennio Salvioni}
\affiliation{Dipartimento di Fisica e Astronomia, Universit\`a di Padova and\\ INFN, Sezione di Padova, Via Marzolo 8, 35131 Padua, Italy}

\author{Bibhushan Shakya}
\affiliation{Deutsches Elektronen-Synchrotron DESY, Notkestr.~85, 22607 Hamburg, Germany}

\begin{abstract}

We present a novel mechanism for thermal dark matter production, characterized by a ``bounce": the dark matter equilibrium distribution transitions from the canonical exponentially falling abundance to an exponentially rising one, resulting in an enhancement of the freezeout abundance by many orders of magnitude. We discuss several realizations of bouncing dark matter. The bounce allows the present day dark matter annihilation cross section to be significantly larger than the canonical thermal target, improving the prospects for indirect detection signals.
  
\end{abstract}

\maketitle

\section{Introduction}

Discovering the underlying nature of dark matter (DM) is one of the main goals of contemporary research in particle physics. Efforts in this direction primarily focus on two key questions: how DM achieved its observed relic abundance, and how its microscopic interactions can be detected with experiments today. DM in thermal equilibrium with the Standard Model (SM) bath in the early Universe follows an abundance distribution that falls exponentially as the Universe cools, until the rates of interactions that keep it in equilibrium become slower than the cosmic expansion rate (see e.g.~\cite{Kolb:1990vq}). This thermal freezeout paradigm represents a strongly motivated and widely studied framework for DM\@.  The simplest realization, known as 
the WIMP (weakly interacting massive particle), makes a generic prediction for the DM annihilation cross section expected today: $\langle \sigma v\rangle_{\text{canonical}} \approx 2\times 10^{-26}$ cm$^3$ s$^{-1}$, providing a compelling target for a variety of current and planned experimental searches for DM\@.

Several variations to this canonical thermal freezeout picture are possible: DM freezeout can be driven by different processes \cite{Griest:1990kh,Carlson:1992fn,Hochberg:2014dra,DAgnolo:2015ujb,Kuflik:2015isi,Kopp:2016yji,Dey:2016qgf,DAgnolo:2017dbv,Berlin:2017ife,DAgnolo:2018wcn,Kim:2019udq,Maity:2019hre,DAgnolo:2020mpt,Kramer:2020sbb,Frumkin:2021zng,Frumkin:2022ror}, can involve interactions with particles whose abundances differ from their equilibrium abundances \cite{Bandyopadhyay:2011qm,Farina:2016llk,Dror:2016rxc,Cline:2017tka}, or feature DM at a temperature different from the temperature of the thermal bath \cite{Feng:2008mu,Fitzpatrick:2020vba}. However, all of these scenarios are still characterized by an exponentially decreasing DM abundance until freezeout. 
Furthermore, in many thermal DM scenarios, the present day annihilation cross section is generally equal to or smaller than $\langle \sigma v\rangle_{\text{canonical}}$, as the existence of stronger interactions would suppress the DM freezeout abundance below its observed value (a few notable exceptions exist; for instance, Sommerfeld enhancement effects \cite{Arkani-Hamed:2008hhe} and dark sectors evolving with a separate temperature and a cannibalistic phase \cite{Pappadopulo:2016pkp}). 

The aim of this paper is to highlight the existence of a novel mechanism for producing thermal DM that deviates from this general pattern. Specifically, we explore scenarios where the DM abundance transitions away from the standard exponentially suppressed distribution to a {\it rising} equilibrium curve in the final stages of freezeout, resulting in an enhancement of the final DM abundance by several orders of magnitude.\,\footnote{This feature has been observed in~\cite{Katz:2020ywn} for metastable dark sector particles (see also~\cite{Griest:1989bq}), then in~\cite{Ho:2022erb,Ho:2022tbw} for DM, but without detailed discussion of the mechanism.} We term this transition a \textit{bounce}, and DM exhibiting such behavior \textit{bouncing dark matter}. A late increase in the DM abundance is possible in various scenarios, e.g.\,\cite{Feng:2003xh,Garny:2017rxs,Forestell:2018dnu}, but out of equilibrium; bouncing dark matter is, to our knowledge, the first realization of this behavior in a thermal context. 

We first provide a technical description of the general conditions necessary for bouncing dark matter (Section~\ref{sec:idea}), followed by a detailed discussion of the physics behind the bounce within a simplified framework (Section~\ref{sec:bounce}). The most salient phenomenological feature of bouncing DM is that the present day DM annihilation cross section can be significantly larger than $\langle \sigma v\rangle_{\text{canonical}}$: while such large cross sections would lead to a too-small relic abundance of DM in standard freezeout scenarios, here the subsequent bouncing phase raises the DM abundance to the correct value. Such enhanced present day annihilation cross sections greatly improve the prospects of discovering DM signals with various indirect detection experiments (Section~\ref{subsec:indirectdetection}). We also present other illustrative examples of bouncing DM (Section~\ref{sec:others}).

\section{Chemistry of the Bounce}
\label{sec:idea}

The evolution of number densities of various species can be tracked via the corresponding chemical potentials $\mu_i$, defined as $n_i \!\approx\!n_i^{\rm eq}e^{\mu_i/T}$, where $n_i^{\rm eq}$ is the  number density of a species in kinetic equilibrium with the photon bath and with vanishing chemical potential. In the $T\!\ll\!m_i$ limit, $n^{\rm eq}_i = g_i \left(\frac{m_i T}{2 \pi}\right)^{3/2}\!e^{-m_i/T}$, where $g_i$ is the number of degrees of freedom of the particle. If an interaction $A_1+ ...+A_p \leftrightarrow B_1+ ...+B_q$ is rapid compared to the expansion rate of the Universe, i.e.~the Hubble parameter, the chemical potentials of the species involved are related as $\mu_{A_1}+ ...+\mu_{A_p}=\mu_{B_1}+ ...+\mu_{B_q}$. The behavior of the chemical potential determines whether a particle undergoes a bounce.

Suppose that DM shares the same chemical potential as some lighter species, $A$, whose abundance $n_A$ does not rise. If $\mu_\chi=\mu_A$, then $n_\chi= (n^{\rm eq}_\chi / n^{\rm eq}_A)\,n_A$, which implies that $n_\chi$ falls exponentially, since $n^{\rm eq}_\chi / n^{\rm eq}_A$ is a falling exponential. Therefore, a departure of the DM chemical potential from the chemical potentials of all lighter states in the thermal bath is a necessary condition for a bounce.

Requiring that the DM comoving number density, or yield, $Y_\chi=n_\chi/s$ (where $s= 2\pi^2 g_{*} T^3/45$ is the total entropy density, and $g_*$ is the effective number of degrees of freedom in the bath) $\textit{rises}$ as the temperature drops imposes a  stringent condition on $\mu_\chi$. Since the yield  scales as $Y_\chi\sim e^{-m_\chi/T} e^{\mu_\chi/T}$, the second exponential must grow faster than the first one drops. More precisely, requiring $Y_\chi$ to rise as the temperature drops, $dY_\chi/dx>0$, where $x \equiv m_\chi/T$, gives
\be
\label{eq:condition}
\mu_\chi (x)+x\frac{d\mu_\chi (x)}{dx}>m_\chi\left(1-\frac{3}{2x}\right)\;,
\ee
in the limits $x\gg 1$ and constant $g_{*}$. Even if $\mu_\chi< m_\chi$, this condition can be satisfied with a sufficiently large $d\mu_\chi/dx$.

We define a state to undergo a bounce if it is in equilibrium, and if its chemical potential satisfies Eq.\,\eqref{eq:condition} at some moment in the early Universe. As discussed above, this requires that the DM chemical potential deviates from those of all other lighter species in the bath.

\section{Bouncing Dark Matter in a Three Particle Framework}
\label{sec:bounce}

We now illustrate the physics behind the bounce within a simplified framework. Consider a dark sector containing three scalar particles -- the DM candidate $\chi$ and two additional states $\phi_1$ and $\phi_2$ -- with the following interactions:
\be
- \mathcal{L} \supset\, \lambda_{\chi 1} \chi^2\phi_1^2+\lambda_{\chi 2} \chi^2\phi_2^2+\lambda_{12} \phi_1^2\phi_2^2+\lambda \phi_2^2\chi\phi_1\,.
\label{eq:interactions}
\ee
The first three terms are couplings between two dark sector species, while the final term represents an interaction involving all three states, which will facilitate the bounce. We present an explicit model that naturally realizes these interactions in Section~\ref{sec:model}. We assume that the dimensionless couplings ($\lambda_i$'s) are comparable in size, $g_i=1$ for all three species for simplicity, and that the particle masses satisfy
\be\label{eq:spectrum}
m_\chi>m_{\phi_2}>m_{\phi_1},\qquad 2\,m_{\phi_2}>m_\chi+m_{\phi_1}\,.
\ee
Furthermore, we assume that all three particles are stable on the timescale over which DM freezeout occurs. This setup contains all the ingredients needed to discuss the general aspects of the bounce mechanism.

\subsection{Physics of the Bounce}

\begin{figure}[t]
\begin{center}
  \includegraphics[width=1.\linewidth]{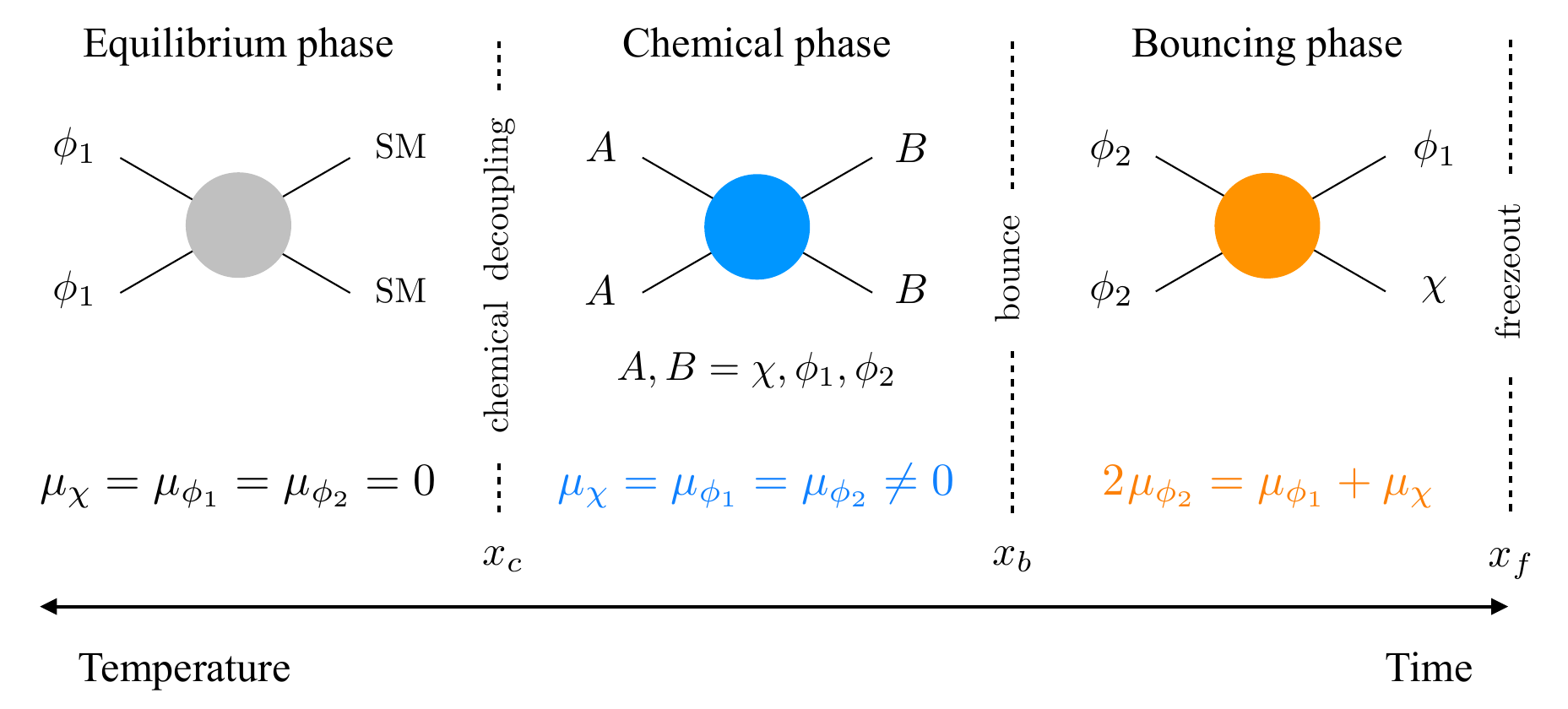}
  \end{center}\vspace{-4mm}
  \caption{A schematic of the three phases in the evolution of dark sector particle abundances, determined by the decoupling of the illustrated processes (processes decouple from left to right). We also show the relations between the chemical potentials of dark sector particles during each phase, which determine their equilibrium abundances.}
\label{fig:schematic}
\end{figure}

A schematic of the decoupling of the processes leading to freezeout of dark sector particles is shown in Fig.\,\ref{fig:schematic}. The cosmological history can be divided into three distinct phases.

In the first (equilibrium) phase, we assume that portal interactions between the dark and SM sectors keep all dark sector particles in thermal equilibrium with the SM bath to temperatures below their masses. Hence all particles follow the standard equilibrium distributions, and 
\be
\mu_\chi=\mu_{\phi_1}=\mu_{\phi_2}=0~~~~\text{(dark $\leftrightarrow$ SM active).}
\ee
We assume that chemical equilibrium between the dark and SM sectors is maintained down to $x=x_c$. Depending on the details of the portal interactions, kinetic equilibrium between the two sectors can last until much later; the implications of this are discussed below.

In the second (chemical) phase, after the dark and SM sectors chemically decouple, the dark species develop nonzero chemical potentials, and the total comoving number density of dark sector particles is conserved.\,\footnote{This requires $4\leftrightarrow 2$ number changing interactions within the dark sector to have decoupled by this point; we have checked that this occurs for our parameters. Note that there are no dark sector $3\leftrightarrow2$ processes, since all the interactions in Eq.\,\eqref{eq:interactions} involve an even number of states.} The evolution of the number densities is now governed by $2\leftrightarrow2$ interactions that can efficiently interconvert the three dark species, $AA\leftrightarrow BB$, where $A,B=\chi, \phi_1,\phi_2$. The chemical potentials thus follow the relations 
\be
\mu_\chi=\mu_{\phi_1}=\mu_{\phi_2}\neq0~~~~\text{(dark $\leftrightarrow$ dark active).}
\label{eq:chem2}
\ee
As $n_i \approx n_i^{\rm eq}e^{\mu_i/T}$, the equilibrium abundances of the heavier states continue to get exponentially suppressed compared to those of the lighter states, with equilibrium distributions uniformly shifted by the common nonzero chemical potential.  

Finally, at $x=x_b$ the system enters the third stage, the bouncing phase, driven by the process $\chi\phi_1\leftrightarrow \phi_2\phi_2$. This occurs when the $AA\leftrightarrow BB$ processes discussed above, as well as the process $\chi\phi_2\leftrightarrow \phi_1\phi_2$, which force the DM to share the same chemical potential as the other dark states, decouple. This imposes a modified relation between the chemical potentials
\be
\mu_\chi+\mu_{\phi_1}=2\mu_{\phi_2}~~~~\text{(only $\chi\phi_1\leftrightarrow \phi_2\phi_2$ active).}
\label{eq:chem3}
\ee 
When $\chi\phi_1\leftrightarrow \phi_2\phi_2$ also decouples, the DM abundance finally freezes out to a constant value.

\begin{figure}[t]
\begin{center}
  \includegraphics[width=\linewidth]{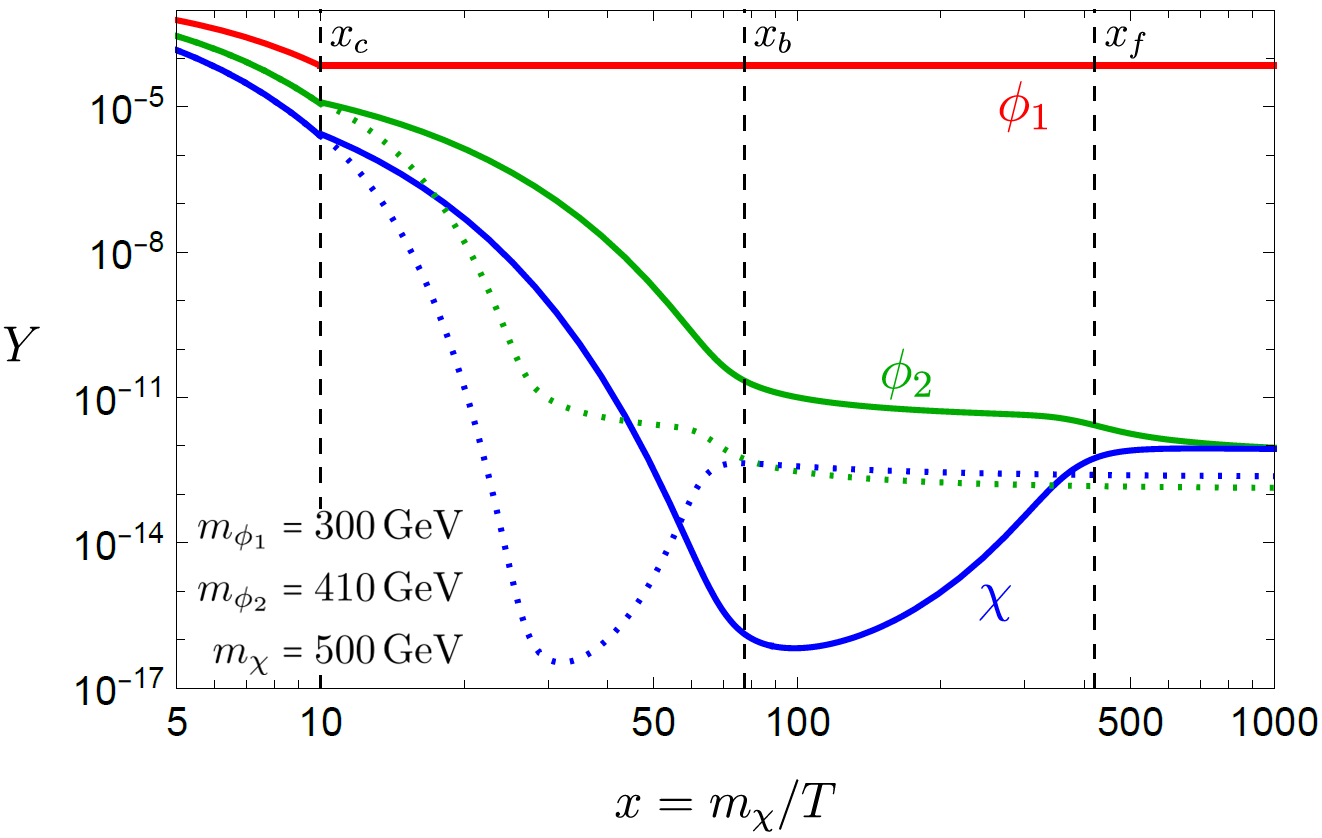}
  \end{center}
  \caption{ Numerical solutions for the yields of dark sector particles for $m_\chi=500$ GeV, $m_{\phi_2}=410$ GeV, and $m_{\phi_1}= 300$ GeV, with cross sections $10\,\sigma_{\chi\chi\phi_2\phi_2}=30\,\sigma_{\phi_2\phi_2\phi_1\phi_1}=10\, \sigma_{\chi\chi\phi_1\phi_1}= \sigma_{\phi_2\phi_2\chi\phi_1}=2.2\times 10^{-25}\,\textup{cm}^3\,\textup{s}^{-1}$, and chemical decoupling at $x_c=10$. The solid curves assume kinetic equilibrium throughout, whereas the dotted curves show the effects of kinetic decoupling of the two sectors at $x_c$. The dashed vertical lines denote the points of (from left to right) chemical decoupling of the dark sector, the bounce, and dark matter freezeout, which separate the various phases discussed in the text, for the kinetically coupled case.}
\label{fig:bounce1}
\end{figure}

\begin{figure}[t]
\begin{center}
  \includegraphics[width=\linewidth]{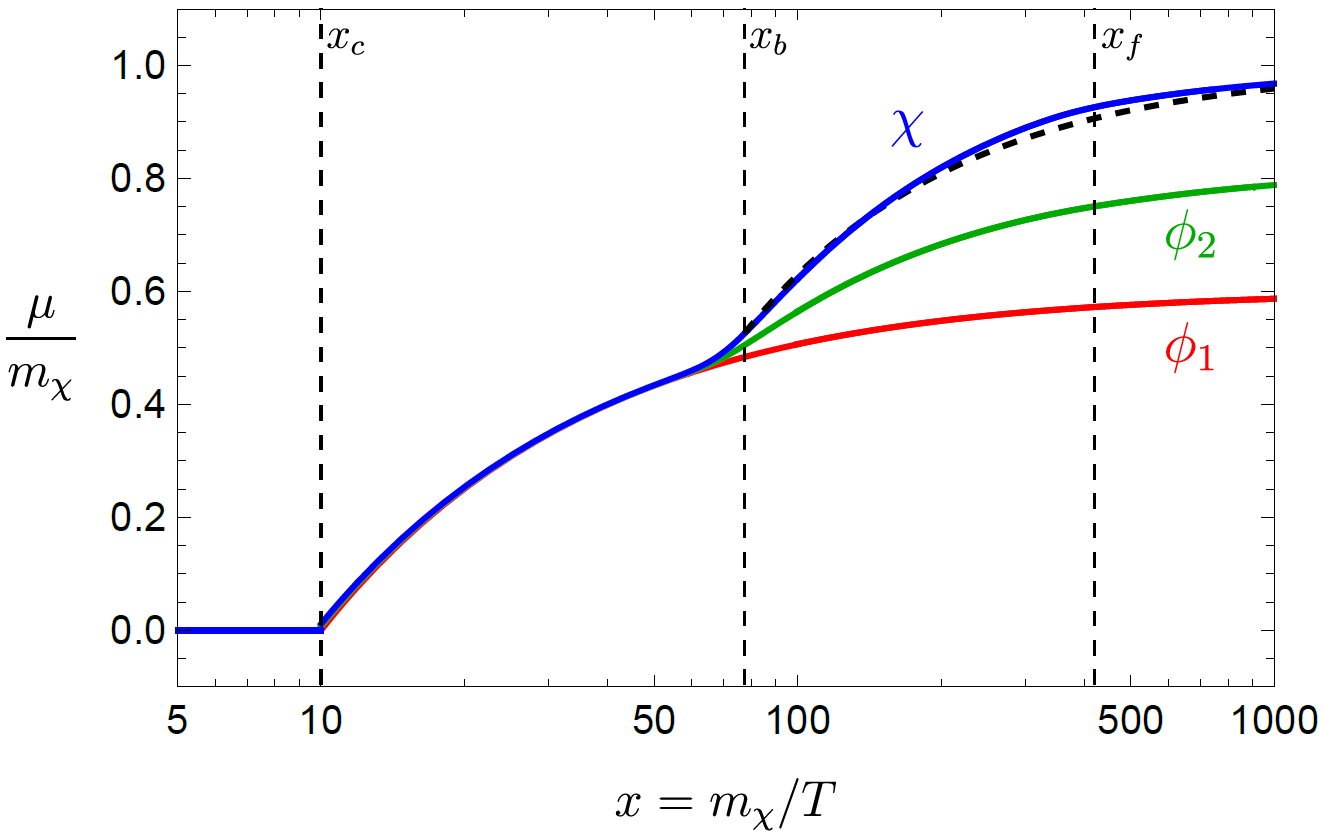}
  \end{center}
  \caption{Numerical solutions for the chemical potentials corresponding to the kinetically coupled case in Fig.\,\ref{fig:bounce1} (solid curves). The dashed black curve corresponds to constant $Y_\chi$ after the bounce.}
\label{fig:chem}
\end{figure}

Figure\;\ref{fig:bounce1} shows the evolution of the yields $Y_i$ for the three dark sector states, obtained by numerically solving the Boltzmann equations for the system (see the Appendix for details) for illustrative benchmark parameters. The solid curves assume kinetic equilibrium throughout, i.e.~all bath particles share a common temperature $T$. The transition between the second and third phases, marked by the ``bounce" from an exponentially falling to an exponentially rising curve for DM, is clearly visible. We show the corresponding evolution of the chemical potentials in Fig.\,\ref{fig:chem}; in particular, note that the bounce corresponds to the instance when the DM chemical potential deviates away from those of the other states.

We also show (dotted curves) the effect of kinetic decoupling between the dark and SM sectors at $x=x_c$, which results in the dark sector cooling faster than the SM bath, with temperature $T_{d}=m_{\chi} x_c/x^2$. This shifts the curves to lower $x$ but otherwise maintains the main qualitative features of the bounce, and the final DM freezeout abundance is only modified by an $\mathcal{O}(1)$ number. Depending on the exact nature of the portal interactions, kinetic decoupling generally occurs at some $x>x_c$, hence the two sets of curves represent the two extremes of late (solid) and early (dashed) kinetic decoupling, and explicit models are expected to fall in between (see also Fig.\,\ref{fig:crosssections}).

The physics behind the bounce in this setup is very intuitive: since $2\,m_{\phi_2}>m_\chi+m_{\phi_1}$, the $ \phi_2\phi_2\to \chi\phi_1$ process is kinematically allowed, whereas the inverse one requires thermal support (note that this is the reverse of standard freezeout dynamics, where processes that deplete DM are kinematically open). Therefore, as the temperature drops, the lighter combination $\chi\phi_1$ is preferentially populated over $\phi_2\phi_2$. Since $\phi_2$ is far more abundant than $\chi$, this results in an exponential \textit{increase} in the comoving number density of $\chi$ as $\phi_2$ particles rapidly get converted into $\chi$ and $\phi_1$.

We now present an analytic discussion of the bounce. The conservation of comoving number density in the dark sector after chemical decoupling can be expressed as
\be
Y_\chi+Y_{\phi_1}+Y_{\phi_2}\equiv Y_S\,,
\label{eq:c1}
\ee
where $Y_S$ denotes the sum of the yields of dark sector particles at the time of chemical decoupling. Next, the relation between chemical potentials after the bounce, Eq.\,\eqref{eq:chem3}, can be rewritten as
\be
Y_{\phi_2}^2=R\, Y_{\phi_1} Y_\chi\,,\qquad R (T)\equiv (n_{\phi_2}^{\rm eq})^2/(n_{\phi_1}^{\rm eq} n_{\chi}^{\rm eq})\,.
\label{eq:c2}
\ee
Furthermore, when $\chi\phi_1\leftrightarrow \phi_2\phi_2$ is the only rapid interaction, $Y_{\phi_1} - Y_{\chi}$ is also conserved, hence 
\be
Y_{\phi_1}-Y_\chi=Y_{\phi_1}^b-Y_\chi^b\equiv Y_D\,,
\label{eq:c3}
\ee
where the superscript $b$ denotes the yields calculated at the bounce. The three equations,~\eqref{eq:c1},~\eqref{eq:c2}, and~\eqref{eq:c3} can be solved analytically for the three unknowns $Y_\chi, Y_{\phi_1},$ and $Y_{\phi_2}$: 
\bea
Y_\chi&=&\frac{4Y_S-(4-R)Y_D-\sqrt{4 R Y_S^2 - (4-R) R Y_D^2}}{2(4-R)}\,,\nonumber\label{eq:dm}\\
Y_{\phi_1}&=&\frac{4Y_S+(4-R)Y_D-\sqrt{4 R Y_S^2 - (4-R) R Y_D^2}}{2(4-R)}\,, \nonumber
\label{eq:Labundance}\\
Y_{\phi_2}&=&\frac{\sqrt{4 R Y_S^2 -(4-R) R Y_D^2} - R Y_S}{4-R}\,.
\label{eq:Mabundance}
\eea
Note that these no longer satisfy $\mu_\chi=\mu_{\phi_1}=\mu_{\phi_2}$. These expressions describe the abundances of the dark sector species in the bouncing phase and are completely determined by the three quantities $Y_S, Y_D$, and $R$. Since $Y_S$ and $Y_D$ are constant, the temperature dependence of the solutions is entirely encoded by
\be
R = \left(\frac{m_{\phi_2}^2}{m_{\phi_1} m_\chi}\right)^{3/2} \exp{\Big(\hspace{-1mm} -\frac{2m_{\phi_2}-m_{\phi_1}-m_\chi}{T}\Big)\,.}
\ee
From this, we see that $2\,m_{\phi_2}<m_\chi+m_{\phi_1}$ and  $2\,m_{\phi_2}>m_\chi+m_{\phi_1}$ lead to drastically different behaviors. In the former case, $R$ rises exponentially as $T$ drops, and $Y_{\chi}$ drops exponentially as a consequence, as is characteristic of standard freezeout processes. However, in the latter case we see the opposite behavior: $R$ falls exponentially with decreasing temperature, hence $Y_\chi$ \textit{increases} after the bounce. 

In the $T \to0$ limit, $R\to 0$, and the DM yield approaches a constant value, 
\be
Y_\chi\to\frac{1}{2}(Y_S-Y_D)=\frac{1}{2}Y_{\phi_2}^b+Y_{\chi}^b\,.
\label{eq:yhlimit}
\ee
In this limit, all of the $\phi_2$ particles present at the bounce are converted to $\chi\phi_1$ at later times, thus contributing the first term, which gets added to the $Y_{\chi}^b$ already present in the bath at the bounce. The enhancement in the dark matter relic abundance relative to the canonical freezeout abundance in this case is
\be
\frac{Y_\chi}{Y_\chi^b}\approx \frac{1}{2} \frac{Y_{\phi_2}^b}{Y_\chi^b} \approx \frac{1}{2}\left(\frac{m_{\phi_2}}{m_\chi}\right)^{3/2} \exp{\Big(\frac{m_\chi-m_{\phi_2}}{T_b}\Big)}\,.
\ee
This ratio is maximized by maximizing the $m_\chi-m_{\phi_2}$ splitting, which occurs for $m_{\phi_1}\to 0$ and $m_\chi\to 2m_{\phi_2}$. Noting that obtaining the correct relic density for weak scale masses in this limit requires $T_b\sim m_{\phi_2}/25$, we estimate that $Y_\chi/Y_\chi^b$ can be as large as $\sim10^{10}$.

In practice, the asymptotic limit in Eq.\,(\ref{eq:yhlimit}) is not reached for two reasons. First, $\chi\phi_1\leftrightarrow \phi_2\phi_2$ freezes out in some finite time. Second, when $Y_\chi\sim Y_{\phi_2}$, the processes $\chi\phi_2\to \phi_1\phi_2$ and/or $\chi\chi \to \phi_2 \phi_2$ become comparable in strength, causing a departure from the above conditions. Consequently, $\chi$ and $\phi_2$ tend to freeze out with comparable abundances. 

$\chi$ and $\phi_2$ can form two-component DM, or $\phi_2$ can decay before Big Bang nucleosynthesis (BBN), leaving only $\chi$ in the late Universe. The very large freezeout abundance of $\phi_1$, however, implies that it must decay before BBN\@. If it gives rise to an early matter dominated era before decaying, the subsequent injection of entropy into the thermal bath will dilute the DM abundance. Decays of dark states are discussed within a concrete model in Section~\ref{sec:model}.

\subsection{Indirect Detection}
\label{subsec:indirectdetection}

In the above setup, the present day DM annihilation cross section $\chi\chi\to \phi_i\phi_i$, which is $s$-wave, has approximately the same size as in the early Universe. If $\phi_i$ decays to SM states, such annihilations can produce observable signals at current and future experiments (see e.g.~\cite{Elor:2015tva,Elor:2015bho,Bell:2016fqf,Barnes:2020vsc,Barnes:2021bsn,Gori:2018lem} for studies of indirect detection signatures from cascade processes in dark sectors). The most interesting phenomenological aspect of bouncing DM is that these cross sections can be significantly larger than $\langle \sigma v\rangle_{\text{canonical}}$. In standard freezeout scenarios driven by DM self-annihilation, increasing $\langle\sigma v\rangle_{\chi\chi\to \phi_i\phi_i}\!>\!\langle \sigma v\rangle_{\text{canonical}}$ would lead to DM tracking its exponentially falling equilibrium curve for longer, freezing out with a relic abundance too small to match observations. For bouncing DM, however, this suppression can be overcome by the exponential enhancement after the bounce, hence larger $\langle\sigma v\rangle_{\chi\chi\to \phi_i\phi_i}$ remains compatible with the observed relic abundance. 

\begin{figure}[t]
\begin{center}
  \includegraphics[width=\linewidth]{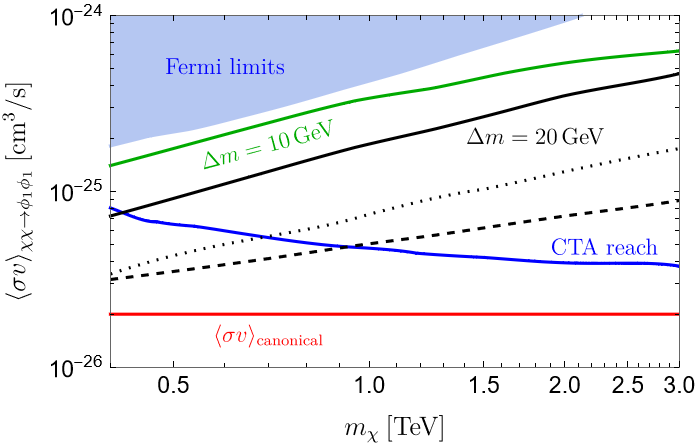}
  \end{center}
  \caption{Predicted present day $\chi\chi\to\phi_1\phi_1$ annihilation cross section as a function of $m_\chi$, with $m_{\phi_1}=m_\chi/2$ and $\Delta m = m_{\phi_2}-(m_\chi+m_{\phi_1})/2=10,\,20$\,GeV (green, black curves respectively). We take $x_c=20$, with the cross sections $40\,\sigma_{\chi\chi\phi_2\phi_2}=200\,\sigma_{\phi_2\phi_2\phi_1\phi_1}=\sigma_{\chi\chi\phi_1\phi_1}=4\, \sigma_{\phi_2\phi_2\chi\phi_1}$, normalized overall to obtain the correct DM relic density for $\chi$. The black dashed and dotted curves correspond, respectively, to the effects of kinetic decoupling of the dark sector at $x=x_c$, or via a Higgs portal (see text for details). The blue shaded region and the solid blue curve denote current bounds from Fermi and projected reach with CTA, respectively, assuming $\phi_1\to WW$. We also show the canonical thermal target $\langle\sigma v\rangle_{\text{canonical}}$ (red line) for reference.}
\label{fig:crosssections}
\end{figure}

This is illustrated in Fig.\,\ref{fig:crosssections}, where we plot the predicted present day $\chi\chi\to\phi_1\phi_1$ annihilation cross section for some representative parameters consistent with the observed DM relic density (we assume $g_*=100$ for simplicity; this affects the cross section results by $\lesssim 15\%$). The nature of the visible signals depends on model-specific details, in particular the dominant decay modes of $\phi_{1}$ and $\phi_2$ (if the latter is unstable).
Here we assume that $\phi_2$ decays away before BBN, so $\chi$ makes up all of DM, and that the $\chi\chi\to \phi_2 \phi_2$ cross section is sufficiently suppressed to be negligible. For concreteness, we also assume the decay mode $\phi_1\to WW$ and choose $m_{\phi_1}=m_\chi/2$, which enables us to adapt results from~\cite{Barnes:2021bsn} to plot bounds from Fermi observations of dwarf galaxies \cite{Fermi-LAT:2015att} and the projected sensitivity from 500 hours of observation of the Galactic Center with the Cherenkov Telescope Array (CTA)~\cite{CTA:2020qlo}.\,\footnote{We use the results from Fig.\,7 of \cite{Barnes:2021bsn} for $\chi\chi\to H'H'$, where $H'$ is a dark Higgs that decays dominantly to $WW$.} We show a baseline case assuming kinetic equilibrium throughout (solid black curve), as well as the modified cross sections to maintain the correct relic density assuming kinetic decoupling at $x_c$ (black dashed), or with the specific choice of a Higgs portal between $\phi_1$ and the SM Higgs doublet $\mathcal{H}$, $\lambda_v \phi_1^2 |\mathcal{H}|^2$ (black dotted), for which the size of $\lambda_v$ controls both chemical and kinetic decoupling (for details on kinetic decoupling calculations, see e.g.~\cite{Kuflik:2015isi}). We thus see that the details of kinetic decoupling can modify the cross section by an $\mathcal{O}(1)$ number. We also show the effects of a smaller mass splitting (solid green), which leads to an enhancement of $n_{\phi_2}$ before the bounce, hence requiring a slightly larger overall cross section to trigger the bounce later and achieve the correct DM relic density. The plot illustrates that the annihilation cross sections for bouncing DM can be larger than the thermal target by more than an order of magnitude (in other parts of parameter space, these cross sections can be much larger or smaller). Note that CTA is unable to reach the thermal target for the chosen decay modes, but can probe bouncing DM for all shown cases over almost the entire mass range, highlighting the improved indirect detection prospects.

\subsection{Model and Constraints}
\label{sec:model}
We now present a concrete realization of the three particle simplified framework. Consider scalar multiplets transforming as $\chi \sim \mathbf{3}_0$, $\phi_2 \sim \mathbf{2}_{+1}$, $\phi_1 \sim \mathbf{1}_0$ under a dark global $SU(2)\times U(1) \times Z_2$ symmetry, with all fields odd under the $Z_2$. This allows the following dark sector interactions at the renormalizable level,
\bea
-\mathcal{L} \supset&\, \lambda_{\chi 1}\mathrm{Tr}(\chi^2) \phi_1^2 + \lambda_{\chi 2}\mathrm{Tr}(\chi^2) |\phi_2|^2 + \lambda_{12} \phi_1^2 |\phi_2|^2\nonumber\\ &+\lambda \phi_2^\dagger \chi \phi_2 \phi_1\,, \qquad \chi \equiv \sigma^a \chi^a\,,
\label{eq:interactions_model}
\eea
where $\sigma^a$ are the Pauli matrices. All other number-changing quartics are automatically forbidden: in particular, $\chi^3\phi_1$ vanishes since $\epsilon^{abc}\chi^a \chi^b \chi^c = 0$, whereas $\chi \phi_1^3$, which would efficiently suppress the bounce, is also not allowed.\,\footnote{The above symmetry structure arises naturally in a three flavor (dark) QCD model with $m_d = m_s$, as considered in~\cite{Katz:2020ywn}.}

In addition, we assume $\phi_1$ couples to the SM as $\mathcal{L} =- \lambda_v \phi_1^2\, \mathcal{H}^\dagger \mathcal{H} + (\bar{g} \phi_1/\Lambda)  W_{\mu\nu}^a W^{a\,\mu\nu}$.\,\footnote{An interesting alternative~\cite{Katz:2020ywn} would be to gauge the dark $U(1)$ and introduce a kinetic mixing of its vector boson with SM hypercharge, thus realizing a vector portal.} The first coupling keeps the dark and visible sectors in equilibrium at early times, while the second coupling (which is just one of many possible choices) explicitly breaks the $Z_2$ and ensures that $\phi_1$ decays to SM particles. As we find below, $\bar{g}$ must be tiny, parametrically smaller than all other couplings in the model. Since all the interactions we consider preserve the dark $SU(2)\times U(1)$, both $\phi_2$ and $\chi$ are stable and can contribute to DM; $\chi$ stability is guaranteed by Eq.\,\eqref{eq:spectrum}, which implies $m_\chi < 2\,m_{\phi_2}$. The presence of two DM components is an interesting feature of this minimal model.

The $\phi_1$ decay width receives two contributions. At tree level, the decays to transverse $WW, ZZ$ give $\Gamma \simeq 3\bar{g}^2 m_{\phi_1}^3/(4\pi \Lambda^2)$. At one loop, a tadpole term $\sim \bar{g}\Lambda^3\phi_1 / (4\pi)^2$ is generated in the scalar potential, leading to a vacuum expectation value (VEV) $\langle \phi_1 \rangle \sim \bar{g}\Lambda^3/[(4\pi)^2 m_{\phi_1}^2]$. As a consequence, $\phi_1$ also decays via the Higgs portal to $hh$ and longitudinal $WW, ZZ$ with $\Gamma \simeq \lambda_v^2 \bar{g}^2 \Lambda^6/ [2\pi (4\pi)^4 m_{\phi_1}^5]$. We require the $\phi_1$ lifetime to be $10^{-6} \lesssim \tau_{\phi_1}/\mathrm{s} \lesssim 1$, i.e.~long enough to enable the bounce, but short enough to not affect BBN\@. Depending on the model parameters, a tighter upper bound on $\tau_{\phi_1}$ can arise if one wishes to avoid an early matter dominated era.

We also require that the trilinear couplings generated by the $\phi_1$ VEV not impact the bounce. These give rise to effective quartics, hence $2\to 2$ processes, that are much smaller than those already present in the Lagrangian provided $\Lambda\,\bar{g}^{1/3}\! \ll\!5\,\mathrm{TeV}\,(m/\mathrm{TeV}) / \lambda_i^{1/6}$, where $m$ is the approximate mass scale of the particles and $\lambda_i$ the generic size of the quartics in Eq.\,\eqref{eq:interactions_model}. The trilinears also give rise to $3\to 2$ processes such as $\phi_1 \phi_1 \phi_1 \to \phi_2 \phi_2$; imposing that these decouple before $x_c$ gives $\Lambda\,\bar{g}^{1/3} \lesssim 2 \,\,\mathrm{TeV}\, (m/\mathrm{TeV})^{7/6} (x_c/10)^{2/3} (10^{-4}/Y_{\phi_1})^{1/3} /\lambda_i^{2/3}$ (using benchmark values from Fig.\,\ref{fig:bounce1}). Finally, requiring that $\phi_1 \phi_1 \to WW$ also decouples before $x_c$ leads to $\Lambda\,\bar{g} \lesssim 0.03\,\,\mathrm{TeV}\, (m/\mathrm{TeV})^{5/4} (x_c/10)^{1/4} (10^{-4}/Y_{\phi_1})^{1/4} /\lambda_i^{1/2}$; notice that for very small $\bar{g}$ this condition is weaker than the previous ones. All the above constraints are satisfied together with those on the $\phi_1$ lifetime in a broad swath of parameter space, spanning $\bar{g} < 10^{-9}$ and $\Lambda > 10$ TeV\@.

Scattering of $\chi$ with nuclei occurs via one-loop processes, with cross-section $\sigma_{\chi N}^{\rm SI} \approx 10^{-48}\,\mathrm{cm}^2 \lambda_{\chi 1}^2 \lambda_v^2\, (\mathrm{TeV}/m_\chi)^2$ (analogous expressions hold for $\phi_2$). For $\lambda_{\chi 1}, \lambda_{1 2} \sim \mathcal{O}(1)$ and $\lambda_v \ll 1$, as is typical in our parameter space, these cross sections are below the neutrino floor. 

\begin{figure}[t]
\begin{center}
  \includegraphics[width=\linewidth]{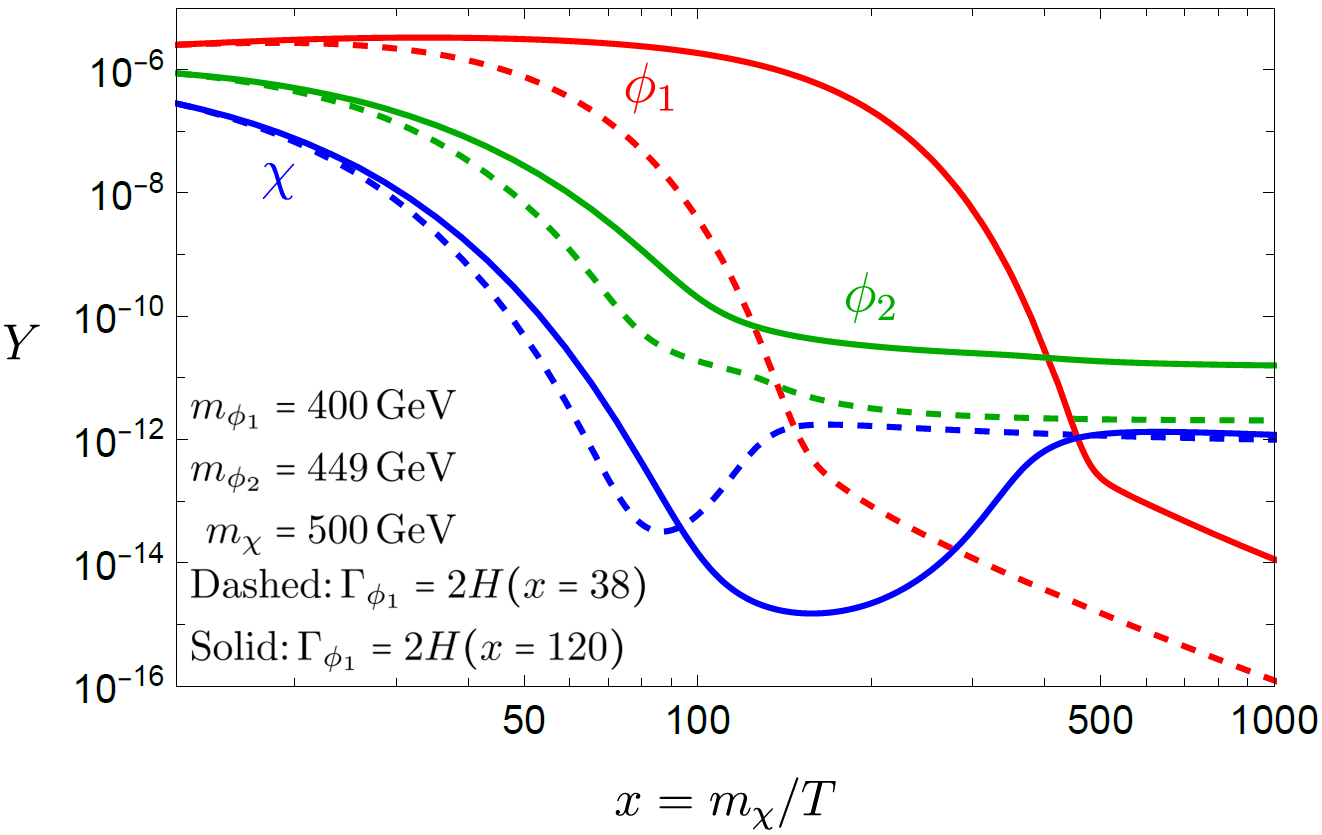}
  \end{center}
  \caption{Numerical solutions for the yields of dark sector particles in the decaying partner scenario (Sec.\,\ref{sec:bounce2}) for $m_{\phi_1}=400\,\textup{GeV}$, $m_{\phi_2}=449\,\textup{GeV}$, $m_{\chi}=500\,\textup{GeV}$, and decaying $\phi_1$, with $x_c=12.5$. The solid, dashed curves correspond to decay widths $\Gamma_{\phi_1}=2H(x=120),\,2H(x=38)$, respectively; note that the latter lifetime is shorter by a factor $10$. The cross sections are chosen to give the correct abundance of $\chi$ DM: $10\,\sigma_{\chi\chi\phi_2\phi_2}=10\,\sigma_{\phi_2\phi_2\phi_1\phi_1}=10\,\sigma_{\chi\chi\phi_1\phi_1}=\sigma_{\phi_2\phi_2\chi\phi_1}=5.5\times 10^{-26}$ (solid), $2.1\times 10^{-26}$(dashed) $\textup{cm}^3\,\textup{s}^{-1}$. }
\label{fig:bounce3}
\end{figure}

Although not required, $\phi_2$ decay can be induced through a $\phi_2$-SM-SM interaction, parametrized by an effective coupling $g_2$ that explicitly breaks the dark global symmetry, leading to $\Gamma_{\phi_2}\! \sim\!g_2^2 m_{\phi_2}/(4\pi)$. $\phi_2$ decays before BBN provided $g_2 \gtrsim 10^{-13}$. This also makes $\chi$ unstable, and we need to ensure that it is sufficiently long-lived to satisfy experimental bounds. If $m_\chi \gtrsim m_{\phi_1} + m_{\phi_2}$, the DM undergoes $4$-body decays with amplitude suppressed by only one insertion of $g_2$, leading to an excessively short lifetime. However, if $m_\chi< m_{\phi_1} + m_{\phi_2}$, DM decays to $5$-body final states (or $3$-body final states via one-loop processes) with $\Gamma_{\chi} \sim \lambda^2 g_2^2\, \mathrm{max}(g_1^2, g_2^2) m_\chi / (4\pi)^7$, where $g_1$ is the effective $\phi_1$-SM-SM coupling that controls $\phi_1$ decay (in the minimal model, $g_1 \sim \bar{g}\, m_{\phi_1}/\Lambda$ if tree level decays dominate). The resulting $\chi$ lifetime satisfies current bounds, yet is potentially interesting for future indirect detection probes~\cite{CTA:2020qlo,Arguelles:2019ouk} of decaying DM: for instance, with $g_1\sim g_2\sim10^{-12}$, $\lambda \sim1$ and $m_\chi\sim\mathrm{TeV}$, we find $\tau_\chi \sim 10^{28}\;\mathrm{s}$.

\section{Other Bounce Scenarios}
\label{sec:others}
In Section \ref{sec:bounce}, we discussed the bounce in the framework of a three particle system with the mass relations in Eq.\,(\ref{eq:spectrum}). However,  bouncing DM can be more broadly realized in other qualitatively different scenarios; it only requires a transition to a new equilibrium curve (such as Eq.\,\eqref{eq:chem3}) that allows the DM chemical potential to depart from those of other species in the bath, and increase sufficiently rapidly to counteract the standard $e^{-m/T}$ suppression. Here, we discuss some other scenarios that realize these conditions. We consider scalar DM for simplicity; however, the bounce can be realized for DM of any spin. For definiteness, kinetic equilibrium is assumed at all temperatures in the examples in this section.

\subsection{Coannihilation with a Decaying Partner}
\label{sec:bounce2}

In Section~\ref{sec:bounce}, the bounce was realized through kinematics; here, we consider a qualitatively different setup that utilizes the decay of a particle to trigger the bounce.

\begin{figure}[t]
\begin{center}
  \includegraphics[width=\linewidth]{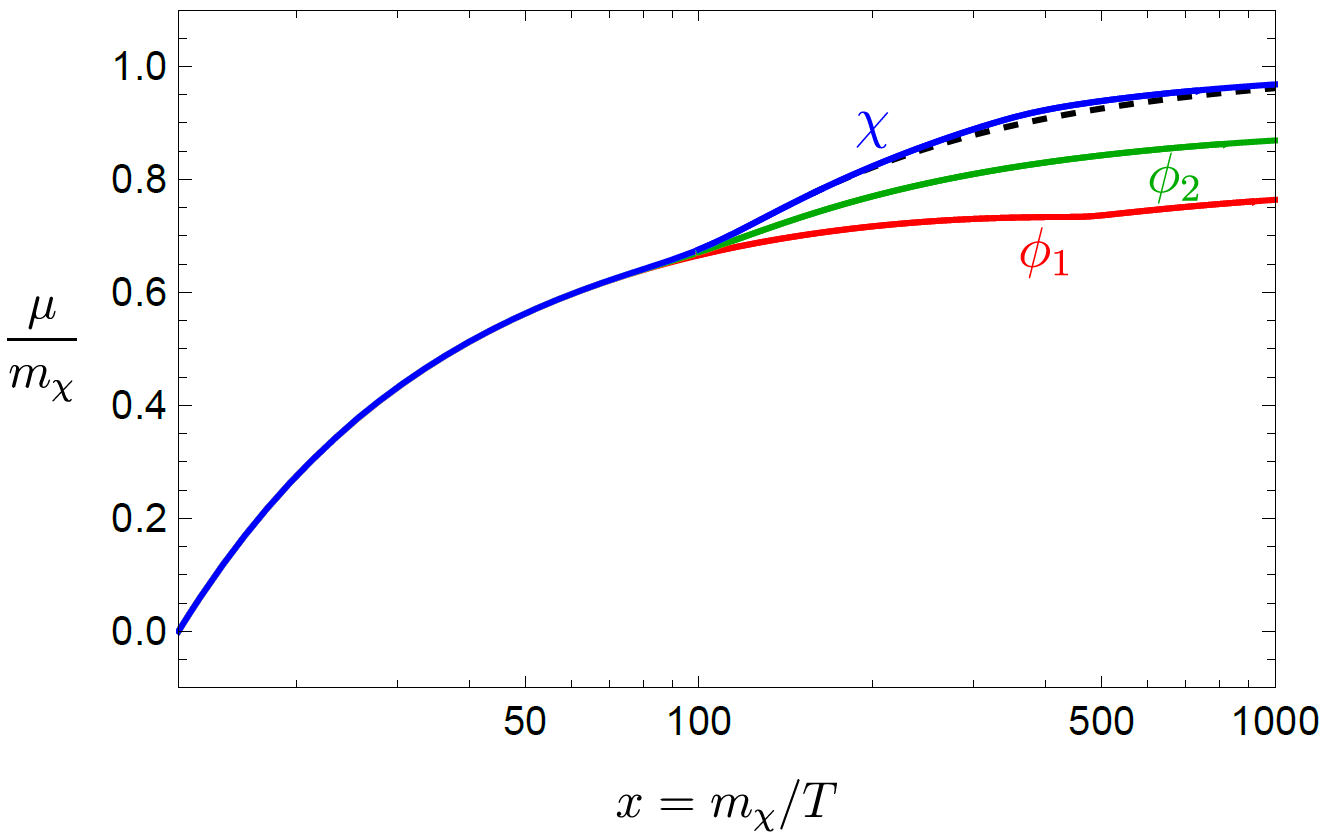}
  \end{center}
  \caption{Numerical solutions for the chemical potentials corresponding to the solid curves in Fig.\,\ref{fig:bounce3}. The dashed black curve corresponds to constant $Y_\chi$ after the bounce.}
  \label{fig:chemdec}
\end{figure}

Consider the same setup as in Section~\ref{sec:bounce}, but with the following modifications:

\vspace{0.5mm}
1. The reversed condition $2\,m_{\phi_2}\lesssim m_\chi+m_{\phi_1}$;\,\footnote{In this case, note that a sufficiently light $\phi_1$ can lead to a rapid decay channel $\chi\to\phi_1\phi_2\phi_2$ even if $\phi_2$ is stable.}

2. $\phi_1$ decays around the time when $\chi$ freezes out, i.e. $\Gamma_{\phi_1}\sim H (x\sim x_f)$.

\vspace{0.5mm}
\noindent Due to the modified relation between the masses, $\phi_2\phi_2\to \chi\phi_1$ is now kinematically closed, and cannot enforce the bounce. Instead, the key ingredient that enables the bounce is the decay of $\phi_1$. To understand this, note that the relation between chemical potentials $\mu_\chi+\mu_{\phi_1}=2\mu_{\phi_2}$ still holds due to $\chi\phi_1\leftrightarrow \phi_2\phi_2$ being rapid in the final stage of freezeout.
The decays of $\phi_1$ cause $\mu_{\phi_1}$ to drop; to maintain the above relation, this must be accompanied by a decrease in $\mu_{\phi_2}$ and an increase in $\mu_\chi$, i.e.~the forward process $\phi_2\phi_2\to \chi\phi_1$ is preferred despite being kinematically disfavored. Since the relations between the yields in Eqs.\,\eqref{eq:c1} and~\eqref{eq:c3} no longer hold due to $\phi_1$ decaying, analytic solutions are difficult to derive. However, the existence of the bounce can be verified numerically, as shown in Fig.\,\ref{fig:bounce3}. The corresponding chemical potentials are shown in Fig.\,\ref{fig:chemdec}.

\subsection{Freezeout Driven by a $3\leftrightarrow 2$ Process }
\label{sec:bounce3}

In contrast to the frameworks considered so far (Section~\ref{sec:bounce} and Section~\ref{sec:bounce2}), we now turn to an example where DM pair-interacts, and the bounce is driven by a $3\leftrightarrow 2$ process. Consider a dark sector with two states, $\chi$ (DM) and $\phi$, with the mass relations
\be
m_\chi>m_\phi\,, \qquad 2 m_\chi<3 m_\phi\,.
\label{eq:m2system}
\ee 
Let us assume that $\chi^2\phi^3$ is the only important interaction (in particular, we assume that $\chi^2\phi^2$ is suppressed and negligible).\,\footnote{At two loops the $\chi^2\phi^3$ interaction can induce $\chi\chi \to \phi\phi$ scattering, which could wash out the bounce~\cite{Ho:2021ojb,Ho:2022erb}. However, the relation between the $\phi\phi\phi\to \chi\chi$ and $\chi\chi \to \phi\phi$ rates is strongly model dependent.} This gives rise to several $3\leftrightarrow2$ number changing interactions, while $2\leftrightarrow2$ processes are absent. When all $3\leftrightarrow2$ interactions such as $\phi\phi\chi\leftrightarrow\phi\chi$ and $\phi\phi\phi\leftrightarrow\chi\chi$ are rapid, the system tracks the standard equilibrium distribution $\mu_\chi=\mu_\phi=0$. As these interactions go out of equilibrium, the system undergoes a bounce at the point where $\phi\phi\phi\leftrightarrow\chi\chi$ remains as the only rapid interaction; this corresponds to a transition to a new equilibrium curve governed by
\be
3\mu_\phi=2\mu_\chi~~~(\phi\phi\phi\leftrightarrow\chi\chi ~\text{active}).
\label{eq:c21}
\ee
This triggers an exponential increase of the DM abundance as $\phi\phi\phi\to \chi\chi$ is kinematically open, while the inverse process requires thermal support. 

This result can be derived analytically by noting that if only $\phi\phi\phi\leftrightarrow\chi\chi$ is active, the following quantity is conserved 
\be
2Y_\phi+3Y_\chi=2Y_\phi^b+3Y_\chi^b\,.
\label{eq:c22}
\ee 
Thus,
\begin{figure}[t]
\begin{center}
  \includegraphics[width=0.95\linewidth, height=5cm]{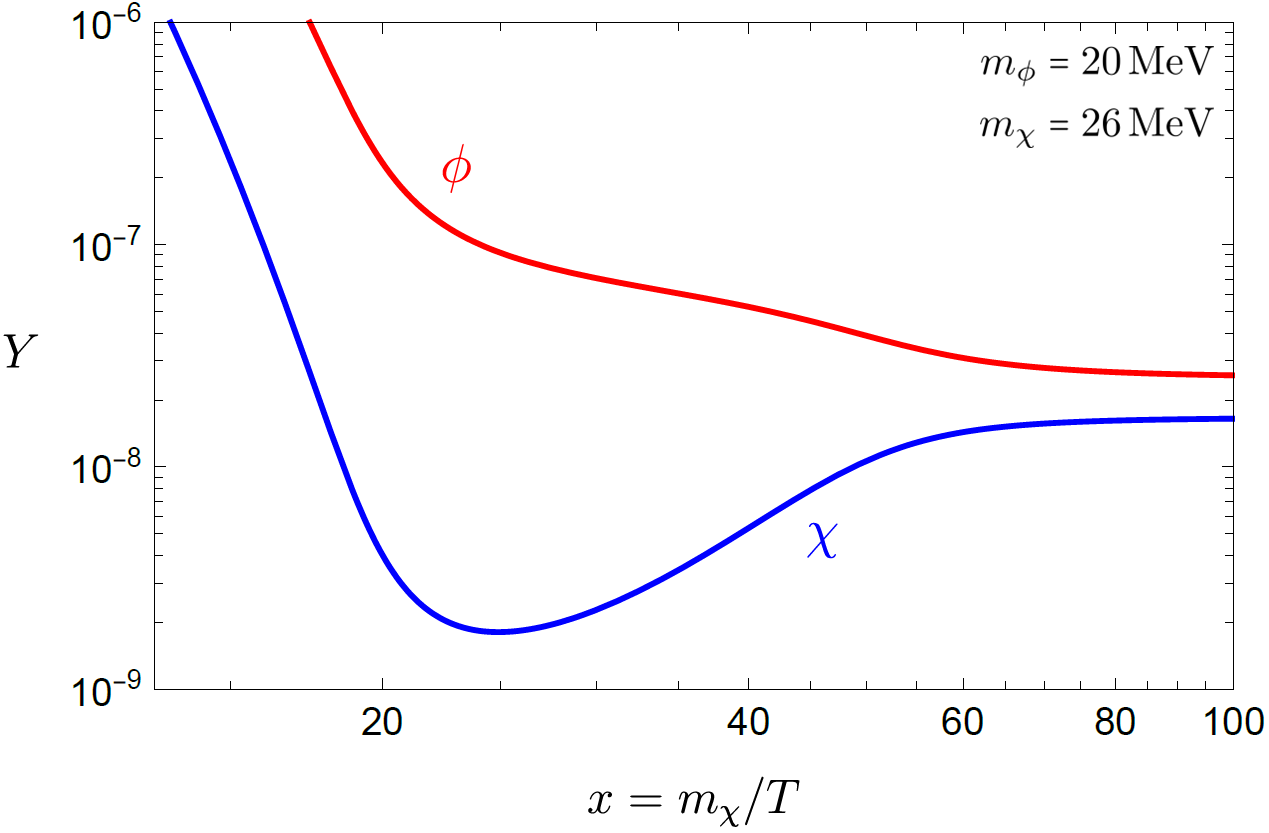}
  \end{center}
  \caption{ Numerical solutions for the yields for $m_\chi=26$ MeV, $m_{\phi}=20$ MeV, and $\sigma_{\phi\phi\phi\chi\chi}=\pi^2 \alpha_{\rm eff}^3/m_\phi^5$ with $\alpha_{\rm eff} = 0.4$. The bounce occurs at $x\sim 20$.}
\label{fig:bounce3to2}
\end{figure}
we have two equations,\,\eqref{eq:c21} and \eqref{eq:c22}, with two unknowns, $Y_\chi$ and $Y_\phi$. Since Eq.\,\eqref{eq:c21} can be rewritten as $(Y_\phi/Y_\phi^{\rm eq})^3=(Y_\chi/Y_\chi^{\rm eq})^2$, we need to solve a cubic equation, and a closed analytic form of the solution, while possible, is not very illuminating. Instead, we note that if the bounce occurs at $T\ll m_\phi, m_\chi$, then $n_\chi\ll n_\phi$, hence in the early stages of the bouncing phase $Y_\phi$ does not decrease appreciably, remaining approximately constant. This implies $\mu_\phi\approx m_\phi-c T$, where the constant $c>0$ is determined by $Y_\phi^b$. Using $Y_\chi\approx e^{-(m_\chi-\mu_\chi)/T}$ together with Eq.\,\eqref{eq:c21}, the $\chi$ yield is
\be
Y_\chi\approx e^{-3 c/2} \,e^{- (2 m_\chi-3m_\phi)/(2T)}\,.
\label{eq:3to2approx}
\ee
This makes it clear that the bounce corresponds precisely to the mass condition $2 m_\chi<3 m_\phi$ in Eq.\,\eqref{eq:m2system}, which leads to an exponential increase of the DM yield. In Fig.\,\ref{fig:bounce3to2}, we show the evolution of the yields for a benchmark case illustrating the bounce in this framework (the DM abundance in the bouncing phase can be approximated by Eq.\,\eqref{eq:3to2approx} with $c\approx 16$). 

The above scenario can arise, for instance, if $\chi$ and $\phi$ are complex scalars whose interactions are mediated only by a heavy scalar $S$ sharing the same quantum numbers as $\chi$. Then the operator $\chi^\ast S |\phi|^2$ is unsuppressed, whereas $\chi^\ast S \phi$ can naturally have a small coupling $\epsilon$ as it violates the $U(1)$ symmetry associated with $\phi\,$-$\,$number. In the effective theory obtained by integrating out $S$, the operator $|\chi|^2|\phi|^2$ is proportional to $\epsilon^2$ and can naturally be much smaller than $|\chi|^2 |\phi|^2 \phi$, which only receives a single $\epsilon\,$-$\,$suppression.

\section{Discussion}

We have introduced the concept of bouncing dark matter, a novel framework for thermal relic dark matter. 
Its defining feature is that DM inherits a large chemical potential, from processes that are in equilibrium,
leading to an exponential rise of the DM abundance before freezeout. This behavior is in stark contrast to most thermal mechanisms, where the DM abundance prior to freezeout is generally characterized by a falling exponential.  
The main phenomenological consequence of the bounce is the possibility of enhanced present day DM annihilation cross sections over the canonical thermal target, which can improve the prospects of DM indirect detection with current and near future experiments. 

In this paper, we have focused on presenting the key physics concepts underlying the bounce within simplified frameworks. It will be interesting to study whether the bounce naturally occurs in existing beyond the Standard Model (BSM) constructions. In general, the bounce requires one or more companion particles with mass comparable to that of DM, which are stable over the timescale of DM freezeout. These conditions are readily realized in extended BSM sectors with nearly degenerate, metastable particles. Such studies can shed light on additional aspects of the bouncing dark matter framework, and therefore represent promising directions for future work.

\section*{Acknowledgments} 

We thank E.~J. Chun, Shu-Yu Ho, Simon Knapen, and Stefan Vogl for helpful discussions and comments. JTR and BS are supported by the Deutsche Forschungsgemeinschaft under Germany’s Excellence Strategy - EXC 2121 Quantum Universe - 390833306. JTR is supported by National Science Foundation grant PHY-2210498. ES acknowledges partial support from the EU’s Horizon 2020 programme under the MSCA grant agreement 860881-HIDDeN\@. JTR, ES, and BS warmly thank the CERN Theory Group, where part of this research was completed, for hospitality. JTR also acknowledges hospitality from the Aspen Center for Physics, which is supported by National Science Foundation grant PHY-1607611. BS also thanks the Berkeley Center for Theoretical Physics for hospitality.

\section*{Appendix}
Here we present the details of the Boltzmann equations that were numerically solved to obtain the abundances of the dark sector particles.

For the three particle framework, we track the yields of the dark sector particles in terms of the dimensionless variable $x=m_{\chi}/T$, where $T$ is the temperature of the SM bath. The initial conditions are set at the time of chemical decoupling of the dark sector, $x=x_c$. In cases where $\phi_1$ decays are not relevant on the bounce timescale (Section \ref{sec:bounce}), $Y_{\phi_1}$ is assumed to be constant after the chemical decoupling. The yields of $\phi_2$ and $\chi$ are obtained by solving the following Boltzmann equations:
\begin{align}
\frac{\mathrm dY_{\phi_2}}{\mathrm dx}
=
-\frac{s(x)}{\tilde{H}(x)x}&\left[2\,\sigma_{\chi\chi\phi_2\phi_2}\left(\frac{(Y_{\chi}^{\textup{eq}})^2}{(Y_{\phi_2}^{\textup{eq}})^2}Y_{\phi_2}^2-Y_{\chi}^2\right)\right.\nonumber\\&\left.+\,2\,\sigma_{\phi_2\phi_2\phi_1\phi_1}\left(Y_{\phi_2}^2-\frac{(Y_{\phi_2}^{\textup{eq}})^2}{(Y_{\phi_1}^{\textup{eq}})^2}Y_{\phi_1}^2\right)\right.\nonumber\\&\left.+\,2\,\sigma_{\phi_2\phi_2\chi\phi_1}\left(Y_{\phi_2}^2-\frac{(Y_{\phi_2}^{\textup{eq}})^2}{Y_{\chi}^{\textup{eq}}Y_{\phi_1}^{\textup{eq}}}Y_{\chi}Y_{\phi_1}\right)\right]\,,
\end{align}

\begin{align}
\frac{\mathrm dY_{\chi}}{\mathrm dx}
=
-\frac{s(x)}{\tilde{H}(x)x}&\left[2\,\sigma_{\chi\chi\phi_2\phi_2}\left(Y_{\chi}^2-\frac{(Y_{\chi}^{\textup{eq}})^2}{(Y_{\phi_2}^{\textup{eq}})^2}Y_{\phi_2}^2\right)\right.\nonumber\\&\left.+\,2\,\sigma_{\chi\chi\phi_1\phi_1}\left(Y_{\chi}^2-\frac{(Y_{\chi}^{\textup{eq}})^2}{(Y_{\phi_1}^{\textup{eq}})^2}Y_{\phi_1}^2\right)\right.\nonumber\\&\left.+\,\sigma_{\phi_2\phi_2\chi\phi_1}\left(\frac{(Y_{\phi_2}^{\textup{eq}})^2}{Y_{\chi}^{\textup{eq}}Y_{\phi_1}^{\textup{eq}}}Y_{\chi}Y_{\phi_1}-Y_{\phi_2}^2\right)\right.\nonumber\\&\left.+\,\sigma_{\phi_2\phi_2\chi\phi_1}\left(Y_{\phi_2}Y_{\chi}-\frac{Y_{\chi}^{\textup{eq}}}{Y_{\phi_1}^{\textup{eq}}}Y_{\phi_2}Y_{\phi_1}\right)\right]\,.
\end{align}
Note that we have replaced the thermally averaged cross sections $\langle \sigma v\rangle_{ABCD}$ with their $T=0$ values $\sigma_{ABCD}$ to lighten the notation, since in this work all processes proceed through the $s$-wave; in general, the proper thermally averaged values should be used. The zero temperature cross sections are related to the quartic couplings in Eq.\,\eqref{eq:interactions} as 
\begin{equation}
\sigma_{AACD} = \tfrac{\lambda_i^2(1 + \delta_{CD})}{8\pi m_A^2}\left( 1 - 2 \tfrac{m_C^2 + m_D^2}{4\,m_A^2} + \tfrac{(m_C^2 - m_D^2)^2}{16\,m_A^4} \right)^{1/2}\,,
\end{equation}
where $\lambda_i$ corresponds to the relevant coupling from Eq.\,\ref{eq:interactions}, and $\delta_{CD}$ is the Kronecker delta. For instance, the benchmark cross sections in Fig.\,\ref{fig:bounce1} correspond to $\lambda_{\chi1} = 0.09, \lambda_{\chi 2} = 0.1, \lambda_{12} = 0.04$, and $\lambda = 0.6$. 

In addition, we have defined $\tilde{H} \equiv H/ [1 + (1/3)\text{d} \log  g_* /\text{d} \log  T]$, where $H=\pi \sqrt{g_*}\, T^2/(3 \sqrt{10}\, M_{\rm Pl})$ is the Hubble parameter, with $M_{\rm Pl}$ the reduced Planck mass, and $s = 2\pi^2 g_\ast T^3 / 45$ is the total entropy density. Recall that $n^{\rm eq}_i = g_i \left(\frac{m_i T}{2 \pi}\right)^{3/2}\!e^{-m_i/T}$ in the nonrelativistic limit $T \ll m_i$. In cases where the dark and SM sectors kinetically decouple at some temperature $T_k$, we solve the above equations with the modified equilibrium distributions corresponding to the modified dark sector temperature, $n^{\rm eq}_i(T)\to n^{\rm eq}_i(T_d)$ with $T_d = T^2/T_k$, which corresponds to instantaneous kinetic decoupling of the dark sector at $T_k$.

When $\phi_1$ decays are relevant for the bounce (Section~\ref{sec:bounce2}), the evolution of the $\phi_1$ abundance is obtained by solving 
\begin{align}
\frac{\mathrm dY_{\phi_1}}{\mathrm dx}
=
-\frac{Y_{\phi_1}\Gamma_{\phi_1}}{\tilde{H}(x)x}&-\frac{s(x)}{\tilde{H}(x)x}\left[2\,\sigma_{\chi\chi\phi_1\phi_1}\left(\frac{(Y_{\chi}^{\textup{eq}})^2}{(Y_{\phi_1}^{\textup{eq}})^2}Y_{\phi_1}^2-Y_{\chi}^2\right)\right.\nonumber\\&\quad\,\left.+\,2\,\sigma_{\phi_2\phi_2\phi_1\phi_1}\left(\frac{(Y_{\phi_2}^{\textup{eq}})^2}{(Y_{\phi_1}^{\textup{eq}})^2}Y_{\phi_1}^2-Y_{\phi_2}^2\right)\right.\nonumber\\&\quad\,\left.+\,\sigma_{\chi\phi_1\phi_2\phi_2}\left(Y_{\chi}Y_{\phi_1}-\frac{Y_{\chi}^{\textup{eq}}Y_{\phi_1}^{\textup{eq}}}{(Y_{\phi_2}^{\textup{eq}})^2}Y_{\phi_2}^2\right)\right.\nonumber\\&\quad\,\left.+\,\sigma_{\chi\phi_1\phi_2\phi_2}\left(\frac{Y_{\chi}^{\textup{eq}}}{Y_{\phi_1}^{\textup{eq}}}Y_{\phi_2}Y_{\phi_1}-Y_{\phi_2}Y_{\chi}\right)\right]~.
\end{align}
The Boltzmann equations for $\phi_2$ and $\chi$ are identical to those shown above, except for the terms corresponding to $\chi_2\chi_2\leftrightarrow\chi\phi_1$, where the $Y^{\rm eq}$ factors need to be appropriately shifted to the other term in the parentheses to reflect the change in mass hierarchy between $2 m_{\phi_2}$ and $m_\chi+m_{\phi_1}$.

For the two particle framework (Section \ref{sec:bounce3}), the relevant Boltzmann equations are
\begin{align}
\frac{\mathrm dY_{\phi}}{\mathrm dx}
&=
-\frac{s(x)^2}{\tilde{H}(x)x}\,\sigma_{\phi\phi\phi\chi\chi}\left[3\left(Y_{\phi}^3-\frac{(Y_{\phi}^{\textup{eq}})^3}{(Y_{\chi}^{\textup{eq}})^2}Y_{\chi}^2\right)\right.\nonumber\\&\quad\,\left.+\left(Y_{\phi}^2Y_{\chi}-Y_{\phi}^{\textup{eq}}Y_{\phi}Y_{\chi}\right)+\left(\frac{(Y_{\chi}^{\textup{eq}})^2}{Y_{\phi}^{\textup{eq}}}Y_{\phi}^2-Y_{\phi}Y_{\chi}^2\right)\right]\,,
\end{align}
\begin{align}
\frac{\mathrm dY_{\chi}}{\mathrm dx}
&=
-\frac{s(x)^2}{\tilde{H}(x)x}\,\sigma_{\phi\phi\phi\chi\chi}\left[2\left(\frac{(Y_{\phi}^{\textup{eq}})^3}{(Y_{\chi}^{\textup{eq}})^2}Y_{\chi}^2-Y_{\phi}^3\right)\right.\nonumber\\&\hspace*{2.9cm} \left.\,+\,2\left(Y_{\phi}Y_{\chi}^2-\frac{(Y_{\chi}^{\textup{eq}})^2}{Y_{\phi}^{\textup{eq}}}Y_{\phi}^2\right)\right]\,.
\end{align}
Here $\sigma_{ABCDE}$ corresponds to the $T\to 0$ limit of $\langle \sigma v^2\rangle_{ABCDE}$, where the thermal average can be evaluated, for example, using the methods in Appendix E of Ref.~\cite{Cline:2017tka}.

\bibliography{DMbounce}{}

\end{document}